\documentclass[12pt, a4paper]{article}

\usepackage{amssymb}
\usepackage{amsthm,color}
\usepackage{latexsym}
\usepackage{graphicx}
\usepackage{fancyhdr}
\usepackage{bbm}
\usepackage{amsmath}
\usepackage{latexsym,amssymb,amsxtra,mathrsfs,times}

\renewcommand{\vec}[1]{\underline{#1}}

\newtheorem{thm}{Theorem}

\newtheorem{lemma}[thm]{Lemma}

\theoremstyle{definition}

\newtheorem{exam}[thm]{Example}

\newcommand{\F}{\mathbb{F}}

\newcommand{\tr}{{\mathrm{Tr}}}
\newcommand{\gf}{{\mathbb F}}
\newcommand{\ca}{{\mathcal{C}_{(d_0,d_1,\cdots,d_t)}^{(1)}}}
\newcommand{\cb}{{\mathcal{C}_{(\widetilde{d}_1,\cdots,\widetilde{d}_t)}^{(2)}}}

\begin{document}

\title{Weight distribution of cyclic codes with arbitrary number of generalized Niho type zeroes}

\author{
Maosheng Xiong\thanks{Department of Mathematics, The Hong Kong University of Science and Technology, Clear Water Bay, Kowloon, Hong
 Kong, China (e-mail: mamsxiong@ust.hk).}, Nian Li\thanks{Department of Mathematics, The Hong Kong University of Science and Technology, Clear Water Bay, Kowloon, Hong
 Kong, China (e-mail: nianli.2010@gmail.com).}, Zhengchun Zhou\thanks{School of Mathematics, Southwest Jiaotong  University, Chengdu, 610031, China (email: zzc@home.swjtu.edu.cn).} and Cunsheng Ding\thanks{Department of Computer Science and Engineering, The Hong Kong University of Science and Technology, Clear Water Bay, Kowloon, Hong Kong, China (email: cding@ust.hk).}}

\maketitle


\begin{abstract}
\footnotesize{
Cyclic codes are an important class of linear codes, whose weight distribution have been extensively studied. Most previous results obtained so far were for cyclic codes with no more than three zeroes. Inspired by the works \cite{Li-Zeng-Hu} and \cite{gegeng2}, we study two families of cyclic codes over $\gf_p$ with arbitrary number of zeroes of generalized Niho type, more precisely $\ca$ (for $p=2$) of $t+1$ zeroes, and $\cb$ (for any prime $p$) of $t$ zeroes for any $t$. We find that the first family has at most $(2t+1)$ non-zero weights, and the second has at most $2t$ non-zero weights. Their weight distribution are also determined in the paper.

\vspace*{2mm}

\vspace*{2mm}
}

\end{abstract}





\section{Introduction}\label{sec-into}

As an important class of linear codes, cyclic codes have been widely used in many areas such as communication and data storage system. Compared with linear codes in general, they have desirable algebraic properties which enable efficient algorithms for encoding and decoding processes. Cyclic codes can also be used to construct other interesting structures such as quantum codes \cite{qc}, frequency hopping sequences \cite{fhs} and so on.

Let $p$ be a prime number and $\mathbb{F}_p$ be the finite field of order $p$. A cyclic code $\mathcal{C}$ of length $n$ over $\mathbb{F}_p$, by the one-to-one correspondence
$$\begin{array}{cccl}
\sigma:& \mathcal{C}&\rightarrow &R:=\mathbb{F}_{p}[x]/(x^n-1)\\
 &(c_0,c_1,\cdots ,c_{n-1})&\mapsto&c_0+c_1x+\cdots +c_{n-1}x^{n-1},
\end{array}$$
can be identified with an ideal of $R$. There exists a unique monic polynomial $g(x)$ with least degree such that $\sigma(\mathcal{C})=g(x)R$ and $g(x)\mid (x^n-1)$. The $g(x)$ is called the \textit{generator polynomial} and $h(x):=(x^n-1)/g(x)$ is called the \textit{parity-check polynomial} of $\mathcal{C}$. For convenience the cyclic code $\mathcal{C}$ is said to have \textit{$t$ zeroes} if $h(x)$ has $t$ irreducible factors over $\mathbb{F}_{p}$. (In literature some authors call that ``the dual of $\mathcal{C}$ has $t$ zeroes'' instead.) $\mathcal{C}$ is called \emph{irreducible} if $t=1$ and \emph{reducible} if $t \ge 2$.

Denote by $A_i$ the number of codewords of $\mathcal{C}$ with Hamming weight $i$, where $0 \le i \le n$. The study of the weight distribution $(A_0,A_1,\cdots ,A_n)$ or equivalently the weight enumerator given by $1+A_1Y+A_2Y^2+\cdots+A_nY^n$ is important in both theory and application, since the weight distribution gives the minimum distance and thus the error correcting capability of the code, and the weight distribution allows the computation of the probability of error detection and correction with respect to some algorithms \cite{Klov}. Moreover, the weight distribution is related to interesting and challenging problems in number theory (\cite{cal,Schroof}).

In recent interesting papers \cite{Li-Zeng-Hu,gegeng2}, in particular in \cite{gegeng2}, the authors obtained the weight distributions of two new classes of cyclic codes with two Niho type zeroes over $\F_p$, one is a binary three-weight code, and the other is a $p$-ary four-weight code for any $p \ge 2$ (here to simplify the notation we consider it the same class for $p=2$ and $p \ge 3$). Moreover, numerical examples showed that some of the codes considered are optimal and have the best known parameters. The purpose of this paper is to extend this work much further in two directions: we obtain the weight distributions of two classes of cyclic codes with \emph{arbitrary number} of \emph{generalized} Niho type zeroes over $\F_p$. More precisely, the binary cyclic code $\ca$ has $t+1$ generalized Niho type zeroes (for any $t \ge 0$) and it is a $(2t+1)$-weight code, and for any $p \ge 2$, the $p$-ary cyclic code $\cb$ has $t$ generalized Niho type zeroes (for any $t \ge 1$) and it is a $2t$-weight code; their weight distributions can also be computed explicitly. The special cases $\ca$ for $t=1$ and $\cb$ for $t=2$ reduce to the work \cite{gegeng2}.

The determination of the weight distribution for a given code in general is an interesting but challenging problem in number theory. In the past decades, the weight distributions of cyclic codes have been studied extensively. Interested readers may refer to \cite{AL06,BM72,BM73,fit,McE74,McE72,Rao10,schmidt,van,Vega1,Vega2,wol} and the survey paper \cite{D-Y12} for irreducible cyclic codes, and to \cite{Ding2,FL08,Feng12,F-M12,holl,luo2,luo3,luo4,Ding1,M09,Mois09,Vega12,Tang12,Xiong1,Xiong2,Xiong3,zeng} and the references therein for cyclic codes with two or three zeroes. However, due to increased difficulties, there are very few results for cyclic codes with more than three zeroes. We mention here the work of Li {\em et al.} \cite{gegeng} who presented a class of reducible cyclic codes with arbitrary number of zeroes and determined its weight distribution by establishing a surprising connection between the involved exponential sums with the spectrum of Hermitian forms graphs, and the works of Yang {\em et al.} \cite{Y-X-D12,Y-X-X14} who obtained the weight distribution of a class of reducible cyclic codes with arbitrary number of zeroes by using Gauss sums and Jacobi sums. Compared with these works, the cyclic codes $\ca$ and $\cb$ seem more interesting because the number of distinct weights of the codes increases very slowly with respect to the number of zeroes. Moreover, the main innovation of the paper, the structured treatment of the Niho exponents enables us to evaluate complicated exponential sums with arbitrary number of Niho type terms -- such techniques are of independent interest and may be useful for other purposes.

This paper is organized as follows. In Section \ref{sec-2} we introduce the cyclic codes $\ca$ and $\cb$ respectively and the main results (Theorems \ref{1:thm1} and \ref{1:thm2}). In Section \ref{sec-3} we prove Theorem \ref{1:thm1}, and in Section \ref{sec-4} we prove Theorem \ref{1:thm2}. The proof of Theorem \ref{1:thm2} depends crucially on the evaluation of a complicated constant $N_r$ which requires special treatments for $p=2$ and $p \ge 3$ individually. To streamline the idea of the proof, we study $N_r$ for $p=2$ in Section \ref{sec-app1} Appendix I and for $p \ge 3$ in Section \ref{sec-app2} Appendix II.

\section{Two families of cyclic codes with generalized Niho type zeroes} \label{sec-2}

We first fix some notation. Let $p$ be a prime and $m$ a positive integer. For simplicity define $q:=p^m$. A positive integer $d$, always understood as modulo $q^2-1$, is called a {\em Niho exponent} if $d \equiv p^j\pmod{q-1}$ for some integer $0 \leq j<2m$. The Niho exponents were originally introduced by Niho \cite{Niho-PhD} who investigated the cross correlation between an $m$-sequence and its decimation. Since then, Niho exponents were further studied and had been used in other research topics. For cyclic codes with two or three zeroes of Niho type, the reader is referred to \cite{Charpin,Li-Zeng-Hu} and the recent work \cite{gegeng2}.

Inspired by the method used to deal with Niho exponents before, we consider exponents of the form $d \equiv \triangle \pmod{q-1}$, where $\gcd(\triangle,q-1)=1$. Note that $d$ is of Niho type if $\triangle=p^j$ for some integer $j$, thus we may call exponents of this form {\em generalized Niho exponents}.


Now $\gf_{q^2}$ is the finite field of order $q^2=p^{2m}$ and $\gamma$ is a generator of the multiplicative group $\mathbb{F}_{q^2}^*:=\mathbb{F}_{q^2} \setminus \{0\}$. We fix some positive integers $h,\triangle$ such that $h \not \equiv 0 \pmod{q+1}$ and $\gcd(\triangle,q-1)=1$.

When $p=2$, for any $t \ge 0$, let $d_0,d_1,\ldots,d_t$ be integers such that
\begin{eqnarray} \label{2:da}
d_j \equiv s_j(q-1)+\triangle \pmod{q^2-1}, \mbox{ where } s_j \equiv j \cdot h+\frac{\triangle}{2} \pmod{q+1}, \forall 0 \le j \le t.
\end{eqnarray}
Here $\frac{1}{2}$ shall be considered as the multiplicative inverse of $2 \pmod{q+1}$ since $q+1$ is odd. We see that $d_j$'s are generalized Niho exponents. \emph{The binary cyclic code $\mathcal{C}_{(d_0,d_1,\cdots,d_t)}^{(1)}$ with $(t+1)$ zeroes $\gamma^{-d_0},\cdots,\gamma^{-d_t}$ consists of elements $c(\vec{a})$ for any $\vec{a}=(a_0,a_1,\ldots,a_t)$ where $a_0 \in \gf_{q}, a_1,\cdots,a_{t}\in\mathbb{F}_{q^2}$, given by}
\begin{equation}\label{2:ca}
\begin{array}{l}
c(\vec{a})=\left(\tr_{q}\left(a_0 \gamma^{d_0 i}\right)+\tr_{q^2}\left(\sum_{j=1}^t a_j \gamma^{d_j i}  \right)\right)_{i=0}^{q^2-2}.
\end{array}\end{equation}
Here $\tr_{q}$ and $\tr_{q^2}$ denote the trace map from $\mathbb{F}_{q}$ and $\mathbb{F}_{q^2}$ to $\mathbb{F}_{p}$ respectively.

When $p$ is a prime (either 2 or any odd prime), let $\widetilde{d}_1,\ldots,\widetilde{d}_t$ be integers such that
\begin{eqnarray} \label{2:db}
\widetilde{d}_j=\widetilde{s}_j(q-1)+\triangle \pmod{q^2-1}, \mbox{ where } \widetilde{s}_j \equiv j \cdot h+\frac{\triangle-h}{2} \pmod{q+1}, \forall 1 \le j \le t.
\end{eqnarray}
This needs some explanation: when $p=2$, then $q \pm 1$ is odd, and $\frac{1}{2}$ shall be considered as a multiplicative inverse of $2 \pmod{q+1}$; however, when $p$ is an odd prime, then $q \pm 1$ is even, hence $\triangle$ is odd, thus for (\ref{2:db}) to make sense it requires that $\triangle \equiv h \equiv 1 \pmod{2}$.

\emph{The $p$-ary cyclic code $\mathcal{C}_{(\widetilde{d}_1,\cdots,\widetilde{d}_t)}^{(2)}$ with $t$ zeroes $\gamma^{-\widetilde{d}_1},\cdots,\gamma^{-\widetilde{d}_t}$ consists of elements $\widetilde{c}(\vec{a})$ for any $\vec{a}=(a_1,\ldots,a_t)$ where $a_1,\cdots,a_{t}\in\mathbb{F}_{q^2}$, given by}
\begin{equation}\label{2:cb}
\begin{array}{l}
\widetilde{c}(\vec{a})=\left(\tr_{q^2}\left(\sum_{j=1}^t a_j \gamma^{\widetilde{d}_j i}  \right)\right)_{i=0}^{q^2-2}.
\end{array}\end{equation}
We have

\begin{thm} \label{1:thm1} Let assumptions be as above. Define
\begin{eqnarray} \label{2:e} e:=\gcd(h,q+1).\end{eqnarray}
\begin{itemize}
\item[(i).] Let $p=2$. Then for any $t$ with $0 \le t \le \frac{q+1}{2e}$, $\ca $ defined by (\ref{2:ca}) is a binary cyclic code of length $(q^2-1)$ and dimension $(2t+1)m$, with at most $(2t+1)$ non-zero weights, each of which is given by
    \[w_j=\frac{q^2-(je-1)q}{2}, \, 0 \le j \le 2t. \]

\item[(ii).] Let $p$ be a prime (either 2 or any odd prime). Then for any $t$ with $1 \le t \le \frac{q+1}{2e}$, $\cb $ defined by (\ref{2:cb}) is a $p$-ary cyclic code of length $(q^2-1)$ and dimension $2tm$, with at most $2t$ non-zero weights, each of which is given by
    \[\widetilde{w}_j=\frac{p-1}{p} \cdot \left(q^2-(je-1)q\right), \, 0 \le j \le 2t-1. \]
\end{itemize}

\end{thm}

The weight distribution of $\ca$ and $\cb$ can be computed. However, to describe the results, we need some notation. First, let $N_0=1,N_1=0$ and define
\begin{eqnarray} \label{2:nr}
N_r:=r! e^r \sum_{\substack{\lambda_2,\lambda_3,\ldots,\\
\sum_{j \ge 2} j \lambda_j=r}} \binom{\frac{q+1}{e}}{\sum_{j } \lambda_j} \left(\sum_{j } \lambda_j\right)! \prod_{j } \frac{\left(B_j/j!\right)^{\lambda_j}}{(\lambda_j)!},\quad \forall \, r \ge 2. \end{eqnarray}
Here the summation is over all non-negative integers $\lambda_2,\lambda_3,\ldots$ such that $\sum_{j \ge 2}j \lambda_j=r$ and
\begin{eqnarray} \label{2:ai} B_j:=q^{-1}(q-1)^j+(-1)^j(1-q^{-1}),\end{eqnarray}
and $\binom{u}{v}$ is the standard binomial coefficient ``$u$-choose-$v$''. It is easy to find that $(q^2-1)|N_r$ for any $r \ge 2$ and $N_2=e(q^2-1),N_3=e^2(q-2)(q^2-1)$, $N_4=e^2(q^2-1)\left\{(e+3)q^2-6eq+6e-3\right\}, \ldots,$ etc.

Next, let $p$ be a prime, and for any $i,j \ge 0$, define $m_{ij}:=\left(jeq-q-1\right)^i$, and define two matrices $M_t^{(1)},M_t^{(2)}$ as \[ M_t^{(1)}=[m_{ij}]_{0 \le i,j \le 2t}, \quad M_t^{(2)}=[m_{ij}]_{0 \le i,j \le 2t-1}\,\,.\]
Since $M_t^{(1)}$ and $M_t^{(2)}$ are Vandermonde matrices of size $(2t+1) \times (2t+1)$ and $2t \times 2t$ respectively, they are both invertible. The weight distribution of $\ca,\cb$ can be described as follows.

\begin{thm} \label{1:thm2} Let assumptions be as in Theorem \ref{1:thm1}.
\begin{itemize}
\item[(i).] Let $p=2$. For $0 \le j \le 2t$, let $\mu_j$ be the frequency of the weight $w_j$ in $\ca$. Let $\vec{\mu}=(\mu_0,\mu_1,\ldots,\mu_{2t})^T$, and let $\vec{b}=(b_0,b_1,\ldots,b_{2t})^T$ where $b_i=q^{2t+1}N_i-\left(q^2-1\right)^i$. Then
    \[\vec{\mu}=\left(M^{(1)}_t\right)^{-1}\vec{b}. \]

\item[(ii).] Let $p$ be a prime (either 2 or any odd prime). For $0 \le j \le 2t-1$, let $\widetilde{\mu}_j$ be the frequency of the weight $\widetilde{w}_j$ in $\cb$. Let $\vec{\widetilde{\mu}}=(\widetilde{\mu}_0,\widetilde{\mu}_1,\ldots,\widetilde{\mu}_{2t-1})^T$, and let $\vec{\widetilde{b}}=(\widetilde{b}_0,\widetilde{b}_1,\ldots,\widetilde{b}_{2t-1})^T$ where $\widetilde{b}_i=q^{2t}N_i-\left(q^2-1\right)^i$. Then
    \[\vec{\widetilde{\mu}}=\left(M^{(2)}_t\right)^{-1}\vec{\widetilde{b}}. \]
\end{itemize}

\end{thm}

Theorems \ref{1:thm1} and \ref{1:thm2} show that the weight distributions of both $\ca$ and $\cb$ can be completely determined for each given $t$. We have the
following comments:
\begin{enumerate}
  \item [1)]  For $\mathcal{C}_{(d_0)}^{(1)}$ and ${\mathcal{C}_{(\widetilde{d}_1)}^{(2)}}$, the results are trivial (see \cite{Niho-PhD});
  \item [2)]  For $\mathcal{C}_{(d_0,d_1)}^{(1)}$ with $\triangle=1$, the result is covered by Theorem 6 in \cite{gegeng2}. However for $\triangle > 1$, the result is new.
  \item [3)] For $\mathcal{C}_{(d_1,d_2)}^{(2)}$ with $\triangle=1$, the results are covered by Theorem 11 for $p=2$ and Theorem 18 for odd prime $p$ in \cite{gegeng2}. If $\triangle > 1$, however, the result is new.
  \item[4)] For $\mathcal{C}_{(d_0,d_1,d_2)}^{(1)}$ with $\triangle=1$ and $h\in\{\frac{1}{2},\frac{1}{4}\}$, the result is covered by \cite{Li-Zeng-Hu}. However for other cases the result is new.
  \item [5)] For any $t\geq 3$, results on $\ca$ and $\cb$ are all new.
\end{enumerate}

Thus, our main results extend previous works \cite{Li-Zeng-Hu} and  \cite{gegeng2} substantially, and present many new cyclic codes of the form $\ca$ and $\cb$ with arbitrary number of zeroes and flexible parameters. 

In what follows, we provide two examples to illustrate the computation of the weight distributions of both $\ca$ and $\cb$.

\begin{exam} Let $p=2$, $2m=8$ $h=2$ and $\triangle=1$, then $q=p^m=16$ and $e=\gcd(h,q+1)=1$. For $t=2$, by \eqref{2:da}, one gets $(s_0,s_1,s_2)=(9,11,13)$ and $(d_0,d_1,d_2)=(136,166,196)$. Theorem \ref{1:thm1} shows $\mathcal{C}_{(136,166,196)}^{(1)}$ has dimension $(2t+1)m=20$ and at most 5 non-zero weights $w_j=\frac{q^2-(je-1)q}{2}, \,0 \le j \le 4$, i.e., $(w_0,w_1,w_2$, $w_3,w_4)=(136,128,120,112,104)$. Moreover, by {\bf Mathematica}, the inverse matrix of $M_t^{(1)}$ with entries $m_{ij}=\left(jeq-q-1\right)^i$ for $0\leq i,j\leq 2t$ is given by
       \[
       (M_2^{(1)})^{-1}=\left(
         \begin{array}{ccccc}
           \frac{-7285}{524288} & \frac{-4807}{393216} & \frac{1267}{786432} & \frac{-23}{393216} & \frac{1}{1572864} \vspace{2mm} \\
           \frac{123845}{131072} & \frac{-5701}{98304} & \frac{-523}{196608} & \frac{19}{98304} & \frac{-1}{393216} \vspace{2mm}\\
           \frac{24769}{262144} & \frac{6225}{65536} & \frac{35}{131072} & \frac{-15}{65536} & \frac{1}{262144} \vspace{2mm}\\
           \frac{-3995}{131072} & \frac{-2909}{98304} & \frac{197}{196608} & \frac{11}{98304} & \frac{-1}{393216} \vspace{2mm}\\
           \frac{2635}{524288} & \frac{1897}{393216} & \frac{-173}{786432} & \frac{-7}{393216} & \frac{1}{1572864} \vspace{2mm}\\
         \end{array}
       \right).
       \]
   On the other hand, one has $N_0=1, N_1=0$, and by \eqref{2:nr} and  \eqref{2:ai}, one can get $N_2=e(q^2-1)=255$, $N_3=e^2(q-2)(q^2-1)=3570$ and $N_4=e^2(q^2-1)\left\{(e+3)q^2-6eq+6e-3\right\}=237405$. This implies $\vec{b}=(b_0,b_1,\ldots,b_{4})^T=(1048575$, $-255, 267321855, 3726834945, 244708934655)^T$. Then, by Theorem \ref{1:thm2}, the weight distribution of $\mathcal{C}_{(136,166,196)}^{(1)}$ is given by
   \[
   \mu=(M_2^{(1)})^{-1}\vec{b}=(353700,377655,250920,30600,35700)^T.
   \]
This is consistent with numerical computation by {\bf Magma} which shows that $\mathcal{C}_{(136,166,196)}^{(1)}$ is a five-weight cyclic code with the weight enumerator
$1+35700Y^{104}+30600Y^{112}+250920Y^{120}+377655Y^{128}+353700Y^{136}$. $\qquad \square$
\end{exam}

While $\cb$ is defined for any prime $p$, we only present an example for an odd prime below.

\begin{exam} Let $p=3$, $2m=4$, $h=3$ and $\triangle=1$, then $q=p^m=9$ and $e=\gcd(h,q+1)=1$. For $t=3$, by \eqref{2:db}, one gets $(\widetilde{s}_1,\widetilde{s}_2,\widetilde{s}_3)=(2,5,8)$ and $(\widetilde{d}_1,\widetilde{d}_2,\widetilde{d}_3)=(17, 41, 65)$. Theorem \ref{1:thm1} shows $\mathcal{C}_{(17, 41, 65)}^{(2)}$ has dimension $2tm=12$ and at most 6 non-zero weights $w_j=\frac{(p-1)(q^2-(je-1)q)}{p}, \,0 \le j \le 5$, i.e., $(w_0,w_1,w_2,w_3,w_4,w_5)=(60, 54, 48, 42, 36, 30)$. Moreover, by {\bf Mathematica}, the inverse matrix of $M_t^{(2)}$ with entries $m_{ij}=\left(jeq-q-1\right)^i$ for $0\leq i,j\leq 2t-1$ is given by
       \[
       (M_3^{(2)})^{-1}=\left(
         \begin{array}{cccccc}
           \frac{-3094}{177147} & \frac{-46357}{3542940} &
      \frac{5695}{1417176} & \frac{-497}{1417176} &
      \frac{17}{1417176} & \frac{-1}{7085880} \vspace{2mm} \\
          \frac{154700}{177147} & \frac{-46675}{354294} & \frac{-667}{177147} &
      \frac{1711}{1417176} & \frac{-19}{354294} & \frac{1}{1417176} \vspace{2mm} \\
\frac{38675}{177147} &
      \frac{37675}{177147} & \frac{-5167}{708588} & \frac{-1099}{708588} &
      \frac{67}{708588} & \frac{-1}{708588} \vspace{2mm} \\
          \frac{-18200}{177147} & \frac{-16525}{177147}& \frac{1852}{177147} & \frac{649}{708588} &
\frac{-29}{354294}   &    \frac{1}{708588}  \vspace{2mm} \\
           \frac{5950}{177147}&
      \frac{21125}{708588}& \frac{-5761}{1417176}& \frac{-361}{1417176}&
      \frac{49}{1417176}& \frac{-1}{1417176}
 \vspace{2mm} \\
           \frac{-884}{177147}& \frac{-7759}{1771470}& \frac{115}{177147}&
\frac{47}{1417176}& \frac{-1}{177147}&
      \frac{1}{708588} \vspace{2mm} \\
         \end{array}
       \right).
       \]
On the other hand, one has $N_0=1, N_1=0$, and by \eqref{2:nr} and  \eqref{2:ai}, one can get $N_2=e(q^2-1)=80$, $N_3=e^2(q-2)(q^2-1)=560$, $N_4=e^2(q^2-1)\left\{(e+3)q^2-6eq+6e-3\right\}=21840$ and $N_5=e^4(q^2-1)(q^2-2q+2)(q-2)+10e^3(q^2-1)(q-1)(q-2)(q+1-e)=439600$. This implies $\vec{b}=(b_0,b_1,\ldots,b_{5})^T=(531440, -80, 42508880, 297094960$, $11565711440, 230344663600)^T$. Then, by Theorem \ref{1:thm2}, the weight distribution of $\mathcal{C}_{(17, 41, 65)}^{(2)}$ is given by
   \[
   \mu=(M_2^{(1)})^{-1}\vec{b}=(163584,205040,113760,40320,6720,2016 )^T.
   \]
This is consistent with numerical computation by {\bf Magma} which shows that $\mathcal{C}_{(17, 41, 65)}^{(2)}$ is a six-weight cyclic code with the weight enumerator $1+2016Y^{30}+6720Y^{36}+40320Y^{42}+113760Y^{48}+205040Y^{54}+163584Y^{60}$. $\qquad \square$

\end{exam}

\section{Proof of Theorem \ref{1:thm1}}\label{sec-3}

Following the notation from Section \ref{sec-2}, let $p$ be a prime (either 2 or any odd prime), $q=p^m$ and let $\gamma$ be a generator of $\mathbb{F}_{q^2}^*$. For any integer $d$, let $h_d(x) \in \gf_p[x]$ be the monic irreducible polynomial of $\gamma^{-d}$ over $\gf_p$. We first prove

\begin{lemma} \label{3:lem1} Let $d=s(q-1)+\triangle$ and let $d'=s'(q-1)+\triangle$ where $s,s',\triangle$ are some integers such that $\gcd(\triangle,q-1)=1$. Then
\begin{itemize}
\item[1)] \[\deg h_d(x)= \left\{\begin{array}{ccl}
m &:& \mbox{ if } \triangle \equiv 2 s \pmod{q+1},\\
2m&:& \mbox{ otherwise.} \end{array}\right.\]

\item[2)] $h_d(x)=h_{d'}(x)$ if and only if $s \equiv s' \pmod{q+1}$ or $s \equiv \triangle - s' \pmod{q+1}$.

\end{itemize}
\end{lemma}
\begin{proof}
For 1), $\deg h_d(x)$ is the least positive integer $k \le 2m$ such that $d \cdot p^k \equiv d \pmod{q^2-1}$. Suppose $k<2m$. Taking the equation modulo $(q-1)$ we have $\triangle p^k \equiv \triangle \pmod{q-1}$. Since $\gcd(\triangle, q-1)=1$, we have $(q-1)|(p^k-1) \Longrightarrow m|k$, hence $k=m$. Hence $dq \equiv d \pmod{q^2-1} \Longrightarrow d = s(q-1)+\triangle \equiv \triangle -2s \equiv 0 \pmod{q+1}$. This proves 1).

As for 2), suppose $h_d(x)=h_{d'}(x)$. This happens if and only if there exists a positive integer $k \le 2m$ such that $d \cdot p^k \equiv d' \pmod{q^2-1}$. Considering the equation modulo $(q-1)$ we find $m|k$, hence $k=m$ or $2m$. If $k=m$, returning to the original equation, we have $qs(q-1)+q \triangle \equiv s'(q-1)+\triangle \pmod{q^2-1} \Longrightarrow s+s' \equiv \triangle \pmod{q+1}$; If $k=2m$, returning to the original equation we find $s(q-1)+\triangle \equiv s'(q-1)+\triangle \pmod{q^2-1} \Longrightarrow s \equiv s' \pmod{q+1}$. This proves 2).
\end{proof}

Now fix any positive integers $h,\triangle$ such that $h \not \equiv 0 \pmod{q+1}$ and $\gcd(\triangle,q-1)=1$. Let $d_0,d_1,\ldots,d_t$ and $\widetilde{d}_1,\ldots,\widetilde{d}_t$ be integers given by (\ref{2:da}) and (\ref{2:db}) respectively. Then we have

\begin{lemma} \label{3:lem2}
\begin{itemize}
\item[1)] Let $p=2$. For any $t$ with $0 \le t < \frac{q+1}{2e}$, we have $\deg h_{d_0}(x)=m$ and $\deg h_{d_i}(x)=2m, \forall \, 1 \le i \le t$. Moreover, $h_{d_i}(x) \ne h_{d_j}(x)$ for any $0 \le i \ne j \le t$.

\item[2)] Let $p$ be a prime. For any $t$ with $1 \le t \le \frac{q+1}{2e}$, we have $\deg h_{\widetilde{d}_i}(x)=2m, \forall \, 1 \le i \le t$. Moreover, $h_{\widetilde{d}_i}(x) \ne h_{\widetilde{d}_j}(x)$ for any $1 \le i \ne j \le t$.
\end{itemize}
\end{lemma}

\begin{proof} Both 1) and 2) follow from Lemma \ref{3:lem1} directly.
\end{proof}

\noindent \emph{Proof of Theorem \ref{1:thm1}.} It implies from Lemma \ref{3:lem2} directly that the cyclic codes $\ca$ and $\cb$ have dimensions $(2t+1)m$ and $2tm$ respectively. To find the number of non-zero weights, we use an idea which is reminiscent of the proof of \cite[Lemma 1]{gegeng2} (see also \cite[Lemma 2]{nh}). We first consider the case $p=2$. For $\ca$, the Hamming weight of a codeword $c(\vec{a})$ can be expressed as
\begin{eqnarray*} \omega_H(c(\vec{a}))&=&q^2-\#\left\{x \in \gf_{q^2}: \tr_q\left(a_0x^{d_0}\right)+\tr_{q^2}\left(\sum_{j=1}^ta_jx^{d_j}\right)=0 \right\}\\
&=&q^2-\frac{1}{2}\sum_{x \in \gf_{q^2}} \sum_{\lambda \in \gf_2}(-1)^{\lambda \tr_q\left( a_0x^{d_0}\right)+\lambda \tr_{q^2}\left( \sum_{j=1}^ta_jx^{d_j}\right)}
=\frac{q^2}{2}-\frac{S(\vec{a})}{2},
\end{eqnarray*}
where
\begin{eqnarray} \label{3:sa} S(\vec{a}):=\sum_{x \in \gf_{q^2}} (-1)^{\tr_q\left(a_0x^{d_0}\right)+\tr_{q^2}\left(\sum_{j=1}^ta_jx^{d_j}\right)}.\end{eqnarray}
We can write it as
\[S(\vec{a})=1+\sum_{x \in \gf_{q^2}^*} (-1)^{\tr_q\left(a_0x^{d_0}+\sum_{j=1}^ta_jx^{d_j}+a_j^q x^{qd_j}\right)}.\]
Define $\bar{x}:=x^q$ for any $x \in \gf_{q^2}$ and $U=\left\{z \in \gf_{q^2}: z \bar{z}=1\right\}$. Then $U$ is a cylic group of order $q+1$. There is a positive integer $\bar{\triangle}$ such that $\bar{\triangle} \triangle \equiv 1 \pmod{q-1}$, and each $x \in \gf_{q^2}^*$ can be written uniquely as $x = y^{\bar{\triangle}}z$ for $y \in \gf_q^*, z \in U$. Such $y$ and $z$ satisfy $y^q=y$ and $\bar{z}=z^q=z^{-1}$. From (\ref{2:da}) we have
\[x^{d_j}=y^{\bar{\triangle}d_j}z^{d_j}=y z^{\triangle-2s_j}=y z^{-2jh}, 0 \le j \le t. \]
Hence
\[S(\vec{a})=1+\sum_{z \in U}\sum_{y \in \gf_{q}^*} (-1)^{\tr_q\left(y\left(a_0 +\sum_{j=1}^ta_jz^{-2jh}+\bar{a}_jz^{2jh}\right)\right)}.\]
Clearly $S(\vec{a})=q(N-1)$, where $N$ is the number of $z \in U$ such that
\[a_0 +\sum_{j=1}^ta_jz^{-2jh}+\bar{a}_jz^{2jh}=0. \]
Let $u=z^{2h}$ and multiply $u^{t}$ on both sides of the above equation, we find
\[a_0u^t+\sum_{j=1}^ta_ju^{t-j}+\bar{a}_j u^{t+j}=0. \]
This is a polynomial of degree at most $2t$, so possibly it may have $0,1,\ldots$, or $2t$ solutions for $u$, and for each valid solution $u$, the number of $z \in U$ such that $z^{2h}=u$ is exactly $e=\gcd(2h,q+1)$. Hence the possible values of $N$ are $je, \forall 0 \le j \le 2t$.  This indicates that $S(\vec{a})$ and $\omega_H(c(\vec{a}))$ take at most $(2t+1)$ distinct values. This proves (i) of Theorem \ref{1:thm1}.

As for $\cb$, if $p=2$, the proof is very similar, so we omit details. 
We only consider that $p \ge 3$. The Hamming weight of a codeword $\widetilde{c}(\vec{a})$ can be expressed as

\begin{eqnarray*} \omega_H(\widetilde{c}(\vec{a}))&=&q^2-\#\left\{x \in \gf_{q^2}: \tr_{q^2}\left(\sum_{j=1}^ta_jx^{\widetilde{d}_j}\right)=0 \right\}\\
&=&q^2-\frac{1}{p}\sum_{x \in \gf_{q^2}} \sum_{\lambda \in \gf_p} \zeta_p^{\lambda \tr_{q^2}\left( \sum_{j=1}^ta_jx^{\widetilde{d}_j}\right)}
=q^2\left(1-\frac{1}{p}\right)-\frac{\widetilde{S}(\vec{a})}{p}\, \, ,
\end{eqnarray*}
where $\zeta_p:=\exp\left(2 \pi \sqrt{-1}/p\right)$ is a complex primitive $p$-th root of unity, and
\begin{eqnarray} \label{3:sb} \widetilde{S}(\vec{a}):=(p-1)+\sum_{\lambda \in \gf_p^*} \sum_{x \in \gf_{q^2}^*} \zeta_p^{\lambda \tr_{q^2}\left(\sum_{j=1}^ta_jx^{\widetilde{d}_j}\right)}.\end{eqnarray}
We write each $x \in \gf_{q^2}^*$ uniquely as $x = y^{\bar{\triangle}} w$ for $y \in \gf_q^*$ and $w \in \Omega:=\{1,\gamma,\gamma^2,\ldots,\gamma^q\}$ (see also \cite[Lemma 2]{nh}) and $\bar{\triangle}$ satisfies $\bar{\triangle} \triangle \equiv 1 \pmod{q-1}$. Then from (\ref{2:db}), we have $\tr_{q^2}(a_jx^{\widetilde{d}_j})=\tr_{q^2}(a_j(y^{\bar{\triangle}}w)^{\widetilde{d}_j})=\tr_{q^2}(a_jyw^{\widetilde{d}_j})=\tr_q\left(a_jyw^{\widetilde{d}_j}+\bar{a}_j y \bar{w}^{\widetilde{d}_j}\right)$,  where $\bar{z}:=z^q$. Hence we obtain
\[\widetilde{S}(\vec{a})=(p-1)\left(1+\sum_{w \in \Omega} \sum_{y \in \gf_q^*} \zeta_p^{\tr_q\left(y\left(\sum_{j=1}^ta_jw^{\widetilde{d}_j}+\bar{a}_j\bar{w}^{\widetilde{d}_j}\right)\right)} \right).\]
Thus $\widetilde{S}(\vec{a})=(p-1)q(N-1)$, where $N$ is the number of $w \in \Omega$ such that
\[\sum_{j=1}^ta_jw^{\widetilde{d}_j}+\bar{a}_j\bar{w}^{\widetilde{d}_j}=0. \]
Dividing $w^{\triangle}$ on both sides, letting $z=w^{q-1}$ and noting that $z^{q+1}=1$, we obtain
\[\sum_{j=1}^t a_jz^{\widetilde{s}_j}+\bar{a}_jz^{\triangle-\widetilde{s}_j}=0. \]
That is,
\[\sum_{j=1}^t a_jz^{j \cdot h+\frac{\triangle-h}{2}}+\bar{a}_jz^{\frac{\triangle+h}{2}-j \cdot h}=0. \]

Multiplying $z^{t \cdot h-\frac{\triangle+h}{2}}$ on both sides of the above equation and setting $u=z^{h}$, then we find
\[\sum_{j=1}^ta_ju^{t+j-1}+\bar{a}_j u^{t-j}=0. \]
This is a polynomial of degree at most $2t-1$, hence it has at most $(2t-1)$ solutions for $u$, and for each valid $u$, the number of $z \in \Omega^{q-1}=U=\{z \in \gf_{q^2}:z^{q+1}=1\}$ such that $z^{h}=u$ is exactly $e=\gcd(h,q+1)$. Hence the possible values of $N$ are $je, \forall 0 \le j \le 2t-1$.  This indicates that $\widetilde{S}(\vec{a})$ and $\omega_H(\widetilde{c}(\vec{a}))$ take at most $2t$ distinct values. This proves (ii) of Theorem \ref{1:thm1} for any odd prime $p$.

Now the proof of Theorem \ref{1:thm1} is complete. \qquad $\square$

\section{Proof of Theorem \ref{1:thm2}}  \label{sec-4}

Since from Theorem \ref{1:thm1} there are only a few non-zero weights in $\ca$ and $\cb$, a standard procedure to determine the weight distribution is to compute power moment identities.

\subsection{For $\ca$}  Here $p=2$. Let $\mu_j$ be the frequency of weight $w_j$ ($0 \le j \le 2t$) in $\ca$. That is,  $\mu_j$ equals the frequency of $\vec{a}=(a_0,a_1,\ldots,a_t)$, $a_0 \in \gf_q,a_1,\ldots,a_t \in \gf_{q^2}$ such that $S(\vec{a})=(je-1)q$, where $S(\vec{a})$ is given by (\ref{3:sa}). Obviously $S(\vec{a})=q^2$ if and only if $\vec{a}=\vec{0}$. We have
\begin{eqnarray} \label{4:id0} q^{1+2t}=1+\sum_{j=0}^{2t} \mu_j, \end{eqnarray}
and for any positive integer $r$,
\begin{eqnarray} \label{4:idr} \sum_{\substack{a_0 \in \gf_q\\
a_j \in \gf_{q^2}, 1 \le j \le t}} \left(S(\vec{a})-1\right)^r=\left(q^{2}-1\right)^r+\sum_{j=0}^{2t}  \left(jeq-q-1\right)^r \mu_j. \end{eqnarray}
On the other hand, by the orthogonal relation
\[\frac{1}{q^2} \sum_{x \in \gf_{q^2}} (-1)^{\tr_{q^2}(xa)}=\left\{\begin{array}{ccl}
0&:& \mbox{ if } a \in \gf_{q^2}^*\\
1&:& \mbox{ if } a =0, \end{array}\right.\]
we find easily that
\begin{eqnarray} \label{4:idr2} \sum_{\substack{a_0 \in \gf_q\\
a_j \in \gf_{q^2}, 1 \le j \le t}} \left(S(\vec{a})-1\right)^r=q^{1+2t}N_r, \end{eqnarray}
where $N_r$ denotes the number of solutions $(x_1,\ldots,x_r) \in \left(\gf_{q^2}^*\right)^r$ to the equations
\begin{eqnarray} \label{4:nra}
\left\{\begin{array}{ccc}
x_1^{d_0}+x_2^{d_0}+\cdots+x_r^{d_0} &=&0, \\
x_1^{d_1}+x_2^{d_1}+\cdots+x_r^{d_1} &=&0, \\
\cdots \cdots & & \\
x_1^{d_t}+x_2^{d_t}+\cdots+x_r^{d_t} &=&0.
\end{array}\right.
\end{eqnarray}
We will prove in Appendix I that for any $0 \le r \le 2t$, $N_r$ is given by the formula (\ref{2:nr}).  Combining identities (\ref{4:id0}), (\ref{4:idr}) and (\ref{4:idr2}) for $1 \le r \le 2t$, we obtain the matrix equation
\[M_t^{(1)} \cdot \vec{\mu}=\vec{b}, \]
where $M_t^{(1)}, \vec{\mu}$ and $\vec{b}$ are explicitly defined before and in Theorem \ref{1:thm2}. Since $M_t^{(1)}$ is invertible, we obtain $\vec{\mu}=\left(M_t^{(i)}\right)^{-1} \cdot \vec{b}$.  Hence (i) of Theorem \ref{1:thm2} is proved.

\subsection{For $\cb$}  Here $p=2$ or any odd prime. Similarly let $\widetilde{\mu}_j$ be the frequency of the weight $\widetilde{w}_j$ ($0 \le j \le 2t-1$) in $\cb$, that is, $\widetilde{\mu}_j$ equals the frequency of $\vec{a}=(a_1,\ldots,a_t)$, $a_1,\ldots,a_t \in \gf_{q^2}$ such that $\widetilde{S}(\vec{a})=(je-1)q(p-1)$ where $\widetilde{S}(\vec{a})$ is given by (\ref{3:sb}). It is known that $\widetilde{S}(\vec{a})=(p-1)q^2$ if and only if $\vec{a}=\vec{0}$. We have
\begin{eqnarray} \label{4:id0b} q^{2t}=1+\sum_{j=0}^{2t-1} \widetilde{\mu}_j, \end{eqnarray}
and for any positive integer $r$,
\begin{eqnarray} \label{4:idrb} \sum_{\substack{
a_j \in \gf_{q^2}, 1 \le j \le t}} \left(\widetilde{S}(\vec{a})-(p-1)\right)^r=(p-1)^r \left(q^{2}-1\right)^r+\sum_{j=0}^{2t-1} (p-1)^r \left(jeq-q-1\right)^r \widetilde{\mu}_j. \end{eqnarray}
On the other hand, using the orthogonal relation
\[\frac{1}{q^2} \sum_{x \in \gf_{q^2}} \zeta_p^{\tr_{q^2}(xa)}=\left\{\begin{array}{ccl}
0&:& \mbox{ if } a \in \gf_{q^2}^*,\\
1&:& \mbox{ if } a =0, \end{array}\right.\]
and noting that as $\lambda$ runs over $\gf_p^*$ once, $\lambda^{\widetilde{d}_j}=\lambda^{\triangle}$ runs $\gf_p^*$ once as well for each $1 \le j \le t$, we obtain
\begin{eqnarray} \label{4:idr2b} \sum_{\substack{a_j \in \gf_{q^2}, 1 \le j \le t}} \left(\widetilde{S}(\vec{a})-(p-1)\right)^r=(p-1)^rq^{2t}N_r, \end{eqnarray}
where $N_r$ denotes the number of solutions $(x_1,\ldots,x_r) \in \left(\gf_{q^2}^*\right)^r$ to the equations
\begin{eqnarray} \label{4:nrb}
\left\{\begin{array}{ccc}
x_1^{\widetilde{d}_1}+x_2^{\widetilde{d}_1}+\cdots+x_r^{\widetilde{d}_1} &=&0, \\
x_1^{\widetilde{d}_2}+x_2^{\widetilde{d}_2}+\cdots+x_r^{\widetilde{d}_2} &=&0, \\
\cdots \cdots && \\
x_1^{\widetilde{d}_t}+x_2^{\widetilde{d}_t}+\cdots+x_r^{\widetilde{d}_t} &=&0.
\end{array}\right.
\end{eqnarray}
Again we will prove in Appendix I for $p=2$ and Appendix II for any odd prime $p$ that for any $0 \le r \le 2t-1$, $N_r$ is also given by the formula (\ref{2:nr}).  Combining identities (\ref{4:id0b}), (\ref{4:idrb}) and (\ref{4:idr2b}) for $1 \le r \le 2t-1$, we obtain the matrix equation
\[M_t^{(2)} \cdot \vec{\widetilde{\mu}}=\vec{\widetilde{b}}, \]
where $M_t^{(2)}, \vec{\widetilde{\mu}}$ and $\vec{\widetilde{b}}$ are explicitly defined before and in Theorem \ref{1:thm2}. We obtain $\vec{\widetilde{\mu}}=\left(M_t^{(2)}\right)^{-1} \cdot \vec{\widetilde{b}}$ and thus (ii) of Theorem \ref{1:thm2} is proved. This completes the proof of Theorem \ref{1:thm2}. \qquad $\square$


\section{Appendix I: Calculation of $N_r$ for $p=2$}\label{sec-app1}

Let us consider $N_r$ for $r \ge 2$. We remark that $N_2,N_3$ were obtained in \cite{gegeng2}, however the computation was somewhat complicated. Here we use a different idea which enables us to compute $N_r$ in general.

\subsection{Case $\ca$} For this case, $N_r$ equals the number of solutions $(x_1,\ldots,x_r) \in \left(\gf_{q^2}^*\right)^r$ to the equations given by (\ref{4:nra}). We write each $x_i \in \gf_{q^2}^*$ as
\[x_i=y_i^{\bar{\triangle}}z_i, \quad y_i \in \gf_q^*, z_i \in U, \]
where $U=\{z \in \gf_{q^2}:z \bar{z}=z^{q+1}=1\}$ is a cyclic group of order $q+1$ and $\bar{\triangle} \triangle \equiv 1 \pmod{q-1}$. Note that this representation of $x_i$ for $y_i \in \gf_q^*, z_i \in U$ is unique.  We obtain
\[x_i^{d_j}=y_i^{\bar{\triangle}d_j}z_i^{d_j}=y_iz_i^{-2jh}, \, 0 \le j \le t. \]
Now denote $u_i=z_i^{-2h} $. Then $u_i \in W:=U^e$, where $e=\gcd(2h,q+1)$. It is clear that $W$ is a cyclic group of order $\frac{q+1}{e}$ and for each $u_i \in W$, there are exactly $e$ many $z_i \in U$ that satisfy the relation $u_i=z_i^{-2h} $. Using these $y_i,u_i$'s, we can write (\ref{4:nra}) as as a matrix equation
\begin{eqnarray}  \label{5:a1}
\left[\begin{array}{cccc}
1&1& \cdots &1 \\
u_1 & u_2 & \cdots &u_r\\
u_1^2 & u_2^2 & \cdots &u_r^2\\
\vdots & \vdots & \ddots &\vdots \\
u_1^t &u_2^t & \cdots &u_r^t
\end{array} \right] \cdot
\left[\begin{array}{c}
y_1\\
y_2\\
\vdots\\
y_r\end{array}\right]=\vec{0},\end{eqnarray}
where we solve for variables $u_i,y_i$ such that $u_i \in W$ and $y_i \in \gf_q^*, 1 \le i \le r$.

Next we take the $q$-th power on both sides of each equation in (\ref{5:a1}) (except the first one). Noting that $y_i^q=y_i$ and $u_i^{q}=u_i^{-1}$, thus we obtain
\begin{eqnarray}  \label{5:a2}
\left[\begin{array}{cccc}
u_1^{-1} & u_2^{-1} & \cdots &u_r^{-1}\\
u_1^{-2} & u_2^{-2} & \cdots &u_r^{-2}\\
\vdots & \vdots & \ddots &\vdots \\
u_1^{-t} &u_2^{-t} & \cdots &u_r^{-t}
\end{array} \right] \cdot
\left[\begin{array}{c}
y_1\\
y_2\\
\vdots\\
y_r\end{array}\right]=\vec{0}. \end{eqnarray}

We combine the matrices in (\ref{5:a1}) and (\ref{5:a2}) together, observing that the exponent of $u_i$ in each column goes consecutively from $-t$ to $t$, this matrix behaves like a Vandermonde matrix whose rank is easy to understand. In particular if $r \le 2t+1$, then the rank of the matrix equals the number of distinct elements in the set $\{u_1,u_2, \ldots,u_r\}$.

\subsection{Partition and type} To compute the number of solutions $u_i \in W,y_i \in \gf_q^*, 1 \le i \le r$ that satisfy both (\ref{5:a1}) and (\ref{5:a2}), we divide the solution set $u_i \in W, y_i \in \gf_q^*$ according to how the elements $u_1,\ldots, u_r$ may match with each other. This matching will be indicated by a partition of the set $\{1,\ldots,r \}$ into a disjoint union of subsets
\[\{1,\ldots,r\}=\bigcup_{\mu=1}^f I_{\mu},\]
which corresponds to valid solutions $u_i,y_i$ such that $u_i=u_j$ if and only if $i,j$ belong to the same set $I_{\mu}$ for some $\mu$.

We will compute the number of solutions $y_i,u_i$ for each such partition. To illustrate the point, let us take an example.

\begin{exam} The partition $\{1,2, \ldots,7\}=\{1,2\} \cup \{3,4\} \cup \{5,6,7\}$ corresponds to the subset of solutions $u_i,y_i$ of (\ref{5:a1}) and (\ref{5:a2}) such that $u_1=u_2=\tau_1,u_3=u_4=\tau_2,u_5=u_6=u_7=\tau_3$, where $\tau_1,\tau_2,\tau_3 \in W$ are distinct, and $y_1,y_2,,\ldots,y_7 \in \gf_q^*$.  Combining this and (\ref{5:a1}), (\ref{5:a2}) we have
\begin{eqnarray} \label{5:exp}
\left[\begin{array}{ccc}
1& \cdots &1 \\
 \tau_1 & \tau_2 &\tau_3\\
 \tau_1^2 & \tau_2^2 & \tau_3^2 \\
 \ddots &\vdots &\vdots \\
\tau_1^t & \tau_2^t & \tau_3^t  \\
 \tau_1^{-1} & \tau_2^{-1} &\tau_3^{-1}\\
 \tau_1^{-2} & \tau_2^{-2} &\tau_3^{-2}\\
 \vdots & \ddots &\vdots \\
\tau_1^{-t} & \tau_2^{-t} &\tau_3^{-t}
\end{array} \right] \cdot
\left[\begin{array}{l}
y_1+y_2\\
y_3+y_4\\
y_5+y_6+y_7
\end{array}\right]=\vec{0}.\end{eqnarray}
The matrix on the left has rank 3, hence we have
\[y_1+y_2=0, \quad y_3+y_4=0, \quad y_5+y_6+y_7=0.  \]
The number of solutions for $y_i \in \gf_q^*, 1 \le i \le 7$ that satisfy the above equations is given by $(q-1)^2 (q-1)(q-2)$. The number of ways to choose $u_i \in W, 1 \le i \le 7$ is given by $3! \binom{(q+1)/e}{3}$. Finally for each $u_i$, $1 \le i \le 7$ there are $e$ ways to choose $z_i \in U$ such that $z_i^{-2h}=u_i$. So the total number of solutions $y_i,z_i$ corresponding to this partition is given by
\[3!(q-1)^2 (q-1)(q-2) \binom{(q+1)/e}{3} e^{7}.  \qquad \square\]
\end{exam}

Now we resume our computation. For a given partition $\cup_{\mu=1}^f I_{\mu}$ of $\{1,\ldots,r \}$, its ``flag'' is defined to be a vector of non-negative integers $\vec{\lambda}=(\lambda_1,\lambda_2,\ldots,)$ where $\lambda_j=\#\{\mu: \#I_{\mu}=j\} $. The previous example has flag $(0,2,1)$, that is $\lambda_1=0,\lambda_2=2,\lambda_3=1$. We make the following observations:
\begin{itemize}
\item[(1)] $\sum_{j} j \lambda_{j}=r$;

\item[(2)] The number of solutions $u_i,y_i$ corresponding to a partition only depends on the flag of the partition;

\item[(3)] The total number of different partitions of $\{1,\ldots,r\}$ for a given flag $\vec{\lambda}$ is (by convention $0!=1$)
\[\frac{r!}{\prod_{j}(\lambda_{j})! (j !)^{\lambda_{j}}}\,\,. \]

\end{itemize}

\subsection{Counting solutions for a partition} Let $\cup_{\mu=1}^f I_{\mu}=\{1,\ldots,r\}$ be a valid partition with flag $\vec{\lambda}$. Similar to the argument in the previous example, since $r \le 2t+1$, the matrix has full rank, we obtain the identity
\begin{eqnarray} \label{5:ai} \sum_{j \in I_{\mu}} y_{j}=0, \quad \forall \, \mu. \end{eqnarray}
Note that if some $I_{\mu}$ contains only one element $j$, this would force $y_j=0$, which is impossible since we require $y_j \ne 0$. So we assume that $\#I_{\mu} \ge 2$ for each $\mu$, or in other words $\lambda_1=0$.

Denote by $B_{\tau}$ the number of solutions $y_1,\ldots,y_{\tau} \in \gf_q^*$ such that
\[y_1+\ldots+y_{\tau}=0. \]
By the inclusion-exclusion principle, it is easy to obtain the formula
\[B_{\tau}=q^{-1}(q-1)^{\tau}+(-1)^{\tau}(1-q^{-1}), \, \forall \tau \ge 2. \]
The number of $y_i$'s that satisfies (\ref{5:ai}) is obviously $\prod_{j \ge 2}(B_j)^{\lambda_j}$. Now we treat $u_i$.

The number of distinct elements in $\{u_1,\ldots,u_r\}$ is $D:=\sum_{j} \lambda_j$, and the number of ways to choose such $u_i$'s for this partition is given by $D! \binom{(q+1)/e}{D}$. On the other hand, each such $u_i$ (there are $r$ of them) gives rise to $e$ many $z_i$'s. In summary, we find that for flag $\vec{\lambda}$, the number of solutions $y_i,z_i$ is given by
\[\frac{r!}{\prod_{j}(\lambda_j)!(j!)^{\lambda_j}} \binom{\frac{q+1}{e}}{D}D! e^{r} \prod_{j} (B_j)^{\lambda_j}. \]
Summing over all $\lambda_j$ such that $\sum_{j \ge 2} j \lambda_j=r$ gives the value $N_r$. This completes the proof of the formula $N_r$ for $\ca$. \qquad $\square$


\subsection{Case $\cb$} Here $N_r$ is the number of solutions $(x_1,\ldots,x_r) \in \left(\gf_{q^2}^*\right)^r$ to the equations given by (\ref{4:nrb}). Using the same notation as before, we write each $x_i$ uniquely as
\[x_i=y_i^{\bar{\triangle}}z_i, \quad y_i \in \gf_q^*, z_i \in U, \]
and we obtain
\[x_i^{\widetilde{d}_j}=y_i^{\bar{\triangle}\widetilde{d}_j}z_i^{\widetilde{d}_j}=y_iz_i^{-(2j-1)h}=y_iu_i^{2j-1}, \, 1 \le j \le t, \]
where $u_i=z_i^{-h} $. Then $u_i \in W:=U^e$, where $e=\gcd(h,q+1)$. $W$ is a cyclic group of order $\frac{q+1}{e}$ and for each $u_i \in W$, there are exactly $e$ many $z_i \in U$ that satisfy the relation $u_i=z_i^{-h} $. Using $y_i,u_i$'s, we can write (\ref{4:nrb}) as a matrix equation
\begin{eqnarray}  \label{5:b1}
\left[\begin{array}{cccc}
u_1 & u_2 & \cdots &u_r\\
u_1^3 & u_2^3 & \cdots &u_r^3\\
\vdots & \vdots & \ddots &\vdots \\
u_1^{2t-1} &u_2^{2t-1} & \cdots &u_r^{2t-1}
\end{array} \right] \cdot
\left[\begin{array}{c}
y_1\\
y_2\\
\vdots\\
y_r\end{array}\right]=\vec{0}. \end{eqnarray}
Again taking the $q$-th power on both sides of each equation we obtain
\begin{eqnarray}  \label{5:b2}
\left[\begin{array}{cccc}
u_1^{-1} & u_2^{-1} & \cdots &u_r^{-1}\\
u_1^{-3} & u_2^{-3} & \cdots &u_r^{-3}\\
\vdots & \vdots & \ddots &\vdots \\
u_1^{-2t+1} &u_2^{-2t+1} & \cdots &u_r^{-2t+1}
\end{array} \right] \cdot
\left[\begin{array}{c}
y_1\\
y_2\\
\vdots\\
y_r\end{array}\right]=\vec{0}. \end{eqnarray}

Combining the matrices in (\ref{5:b1}) and (\ref{5:b2}) together and noting that the exponent of $u_i$ in each column goes consecutively from $-2t+1$ to $2t-1$ with gap 2, we see that this matrix also behaves like a Vandermonde matrix whose rank is easy to understand. In particular if $r \le 2t$, then the rank of the matrix equals the number of distinct elements in the set $\{u_1,u_2, \ldots,u_r\}$.  This is the only property which was used in the argument for the previous case $\ca$. Hence we conclude that $N_r$ could be computed in exactly the same way as before, and it is given by the formula (\ref{2:nr}) for $r \le 2t$. This completes the proof for the case $\cb$. \qquad $\square$


\section{Appendix II: Calculation of $N_r$ for $p$ odd}\label{sec-app2}

Now $p$ is an odd prime, $q=p^m$ and $N_r$ is the number of solutions $(x_1,\ldots,x_r) \in \left(\gf_{q^2}^*\right)^r$ to the equations given by (\ref{4:nrb}). Let $\gamma$ be a generator of $\gf_{q^2}^*$. Using the same notation as before, we may write each $x_i \in \gf_{q^2}^*$ as
\begin{eqnarray} \label{6:xi} x_i=y_i^{\bar{\triangle}}z_i \epsilon_i, \quad y_i \in \gf_q^*, z_i \in U, \epsilon_i \in \{\gamma,1\}. \end{eqnarray}
Since
\[\gf_q^* \bigcap U=\{1,-1\}, \]
as $y_i,z_i,\epsilon_i$ run over the sets $\gf_q^*,U$ and $\{\gamma,1\}$ once respectively, the $x_i$ will run over $\gf_{q^2}^*$ exactly twice.  So $N_r=2^{-r}M_r$ where $M_r$ is the number of $y_i \in \gf_q^*,z_i \in U,\epsilon_i \in \{\gamma,1\}, 1 \le i \le r$ such that the $x_i$'s from (\ref{6:xi}) satisfy the equations (\ref{4:nrb}) simultaneously. We obtain

\[x_i^{\widetilde{d}_j}=y_i^{\bar{\triangle}\widetilde{d}_j}z_i^{\widetilde{d}_j}\epsilon_i^{\widetilde{d}_j}=y_iz_i^{-(2j-1)h}\epsilon_i^{\widetilde{d}_j}=y_i\left(z_i\epsilon_i^{-(q-1)/2}\right)^{-(2j-1)h}\epsilon_i^{\triangle (q+1)/2}, \, 1 \le j \le t. \]
Moreover, since $\epsilon_i^{\widetilde{s}_j(q-1)} \in U$,
\[x_i^{q\widetilde{d}_j}=y_iz_i^{(2j-1)h}\epsilon_i^{q\widetilde{d}_j}=y_i\left(z_i\epsilon_i^{-(q-1)/2}\right)^{(2j-1)h}\epsilon_i^{\triangle (q+1)/2}, \, 1 \le j \le t. \]

Define $u_i=\left(z_i\epsilon_i^{-(q-1)/2}\right)^{-h} , \xi_i=\epsilon_i^{\triangle(q+1)/2}$. Then $u_i\epsilon_i^{(q-1)h/2} \in W:=U^e$, where $e=\gcd(h,q+1)$. $W$ is a cyclic group of order $\frac{q+1}{e}$ and for each $u_i\epsilon_i^{(q-1)h/2} \in W$, there are exactly $e$ many $z_i \in U$ that satisfy the relation $u_i=\left(z_i\epsilon_i^{-(q-1)/2}\right)^{-h}$. Using $y_i,u_i,\xi_i$'s, we can write (\ref{4:nrb}) as a matrix equation
\begin{eqnarray}  \label{6:b1}
\left[\begin{array}{rrrr}
u_1\xi_1 & u_2 \xi_2 & \cdots &u_r \xi_r\\
u_1^3 \xi_1 & u_2^3 \xi_2& \cdots &u_r^3 \xi_r\\
\vdots & \vdots & \ddots &\vdots \\
u_1^{2t-1} \xi_1 &u_2^{2t-1} \xi_2 & \cdots &u_r^{2t-1} \xi_r
\end{array} \right] \cdot
\left[\begin{array}{c}
y_1\\
y_2\\
\vdots\\
y_r\end{array}\right]=\vec{0}. \end{eqnarray}
From the $q$-th power of each equation we obtain
\begin{eqnarray}  \label{6:b2}
\left[\begin{array}{rrrr}
u_1^{-1} \xi_1& u_2^{-1} \xi_2& \cdots &u_r^{-1} \xi_r\\
u_1^{-3} \xi_1& u_2^{-3} \xi_2& \cdots &u_r^{-3} \xi_r\\
\vdots & \vdots & \ddots &\vdots \\
u_1^{-2t+1} \xi_1 &u_2^{-2t+1} \xi_2& \cdots &u_r^{-2t+1} \xi_r
\end{array} \right] \cdot
\left[\begin{array}{c}
y_1\\
y_2\\
\vdots\\
y_r\end{array}\right]=\vec{0}.\end{eqnarray}

Combining the matrices in (\ref{6:b1}) and (\ref{6:b2}) together and noting that each  column is a multiple of $\xi_i$ and the exponent of $u_i$ goes consecutively from $-2t+1$ to $2t-1$ with gap 2, hence this matrix also behaves like a Vandermonde matrix whose rank is easy to understand. In particular if $r \le 2t$, then the rank of the matrix equals the number of distinct elements in the set $\{u_1,u_2, \ldots,u_r\}$.  Using this property, we find that for each fixed $\vec{\xi}=(\xi_1, \ldots,\xi_r)$, the number of solutions $u_i,y_i$ that satisfy (\ref{6:b1}) and (\ref{6:b2}) is also given by the formula (\ref{2:nr}). On the other hand, each $\xi \in \{\gamma^{\triangle(q+1)/2},1\}$ can take two distinct values, hence $\vec{\xi}$ takes $2^r$ distinct values. Taking into account that $N_r=2^{-r}M_r$, we find that this $N_r$ is exactly the same as given by the formula (\ref{2:nr}). This completes the proof for the case $\cb$ when $p$ is odd. \qquad $\square$

\section{Conclusions}\label{sec-conclusion}
In this paper, for any prime $p$, we determined the weight distributions of two families of cyclic codes over $\gf_p$ with arbitrary number of zeroes of generalized Niho type, more precisely the cyclic codes $\ca$ (for $p=2$) of $t+1$ zeroes given by (\ref{2:ca}) have at most $(2t+1)$ non-zero weights, and the cyclic codes $\cb$ (for any prime $p$) of $t$ zeroes given by (\ref{2:cb}) have at most $2t$ non-zero weights. 

\section*{Acknowledgement}
M. Xiong's research was supported by the Hong Kong Research Grants Council under Grant Nos. 609513 and 606211. Z. Zhou's research was supported by the Natural Science Foundation of China under Grant No. 61201243, and also the Application Fundamental Research Plan Project of Sichuan Province under Grant No. 2013JY0167.  C. Ding's research was supported by The Hong Kong Research
Grants Council under Grant No. 600812.

\end{document}